\documentclass[pra,twocolumn,showpacs]{revtex4}

\usepackage{amssymb}
\usepackage{amsfonts}
\usepackage{amsmath}
\usepackage{epsfig}

\newcommand{\Xstate}{\mbox{X$^1\Sigma_{\mathrm{g}}^+$}}
\newcommand{\astate}{\mbox{a$^3\Sigma_{\mathrm{u}}^+$}}
\newcommand{\rcm}{\mbox{cm$^{-1}$}}
\newcommand{\wn}{$\mathrm{cm}^{-1}$}

\begin{document}

\title{Spectroscopy of the a$^{3}\Sigma_u^{+}$ state and the coupling to the X$^{1}\Sigma_g^{+}$ state of K$_2$}

\author{A.~Pashov}
\affiliation{Department of Physics, Sofia University, 5 James Bourchier boulevard, 1164 Sofia, Bulgaria}

\author{P.~Popov}
\affiliation{Department of Physics, Technical University of Varna, ul. Studentska 2, 9010 Varna, Bulgaria}

\author{H.~Kn\"ockel}
\author{E.~Tiemann}
\affiliation{Institut f\"ur Quantenoptik, Leibniz Universit\"at Hannover, Welfengarten 1, 30167 Hannover,
Germany}


\date{Received: date / Revised version: date}

\begin{abstract}
We report on high resolution Fourier-transform spectroscopy of fluorescence to the \astate~state induced by two-photon or two-step excitation from the \Xstate\ state to the $2^3\Pi_g$\ state in the molecule K$_2$. These spectroscopic data are combined with recent results of Feshbach resonances and two-color photoassociation spectra for deriving the potential curves of \Xstate~and \astate~up to the asymptote. The precise relative position of the triplet levels with respect to the singlet levels was achieved by including the excitation energies from the \Xstate\ state to the $2^3\Pi_g$ state and the frequencies of the fluorescence down to the \astate\ state in the simultaneous fit of both potentials. The derived precise potential curves allow for reliable modeling of cold collisions of pairs of potassium atoms in their $^2S$ ground state. 
\end{abstract}
\pacs{31.50.Bc, 33.20.Kf, 33.20.Vq, 33.50.Dq}

\maketitle


\section{Introduction}
\label{intro}

Contrary to the heteronuclear alkali diatomic molecules (e.g.
\cite{Pashov:05,Docenko:06,Staanum:07}), the lowest triplet state
a$^3\Sigma^+_{u}$ of the homonuclear ones is much less accurately
characterized.  
The experimental data in this case are either fragmentary
or from low resolution spectroscopy. The situation can be understood mainly by
the presence of the gerade/ungerade symmetry in the homonuclear
diatomics which makes the spectroscopic techniques with
single-photon excitation inapplicable. On the other hand the
demand for accurate data on both states, \Xstate\ and \astate, 
correlated to the lowest s + s asymptote of the alkalies, is 
high because of the very active research in the field of ultracold 
collisions on alkali species.

The first spectroscopic observation of the \astate\ state in K$_2$ with
partially resolved rotational structure was reported in Ref.~\cite{Li:90}.
There, blue fluorescence to the \astate\ state was induced with the 
optical-optical double resonance (OODR) technique and resolved with a 0.85~m
dual grating monochromator. The highest observed vibrational level of the 
ground triplet state in $^{39}$K$_2$ was v~=~17.
In a further paper \cite{Zhao:96} the same group reported additional 
OODR measurements on the \astate\ state in order to resolve the problem 
that the derived potential curve of the \astate\ state crossed that of the
\Xstate\ state taken from Ref.~\cite{Amiot:95}. The paper contains a few term energies of low
vibrational levels of the \astate\ state with other rotational 
quantum numbers than the levels observed in \cite{Li:90}. 

The lowest atomic asymptote of K$_2$ was studied by Wang et al~\cite{Wang:00} through two-color photoassociation 
spectroscopy of ultracold $^{39}$K atoms. In the range between 1500 and 4600 MHz below the asymptote a total of 12 term energies of
near asymptotic levels with high triplet character were determined. Thus the analysis of the
\astate\ state level structure presented in these three
publications \cite{Li:90,Zhao:96,Wang:00} could be performed as single channel cases ignoring the singlet-triplet mixing due to
the hyperfine interactions with the \Xstate\ state.

In another group of papers \cite{Loftus:02,Regal:03,Ticknor:04,Regal:04,Gaebler:07} 
s- and p-wave Feshbach resonances in $^{40}$K were measured and just recently 
s-wave Feshbach resonances in $^{39}$K were reported \cite{Errico:07} followed by an application
for successful Bose-Einstein condensation of  $^{39}$K \cite{Roati:07}. Bose-Einstein condensation 
was observed earlier for the isotope $^{41}$K by Modugno et al. \cite{Modugno:01} leading to an 
independent estimate of the triplet scattering length of that isotope.

In their recent work Chu et al. \cite{Chu:05} examined series of
two-step two-color and two-photon single-color laser excitations in K$_2$ which were
devoted to a study of its 2$^{3}\Pi_{g}$ state. Along with the
main subject of their study, the authors observed also laser
induced fluorescence to the triplet a$^3\Sigma^+_{u}$ state.
Unfortunately, this was done at low resolution and thus gave no
additional spectroscopic data for a reliable characterization of
the a$^3\Sigma^+_{u}$ state.

The purpose of our present study is to record high resolution spectra
for the lowest triplet state in K$_2$ and to construct potential 
energy curves accurate enough to model cold collisions between two 
potassium atoms in the coupled system of \Xstate\ and \astate\ states. We will 
investigate the importance of the singlet-triplet mixing  
for relatively deeply bound and asymptotic vibrational levels, since our experience on the heavier 
alkali compounds has shown that by ignoring it one is not able to reproduce 
satisfactorily the whole set of experimental observations (see e.g. \cite{Pashov:05}). 
Finally, the presence of accurate experimental data for several potassium isotopes gives the opportunity
to look for the possible breakdown of the Born-Oppenheimer approximation and,  
consequently, the widely used mass-scaling for cold collisions.

\section{Experiment}

\begin{figure*}
  \centering
\epsfig{file=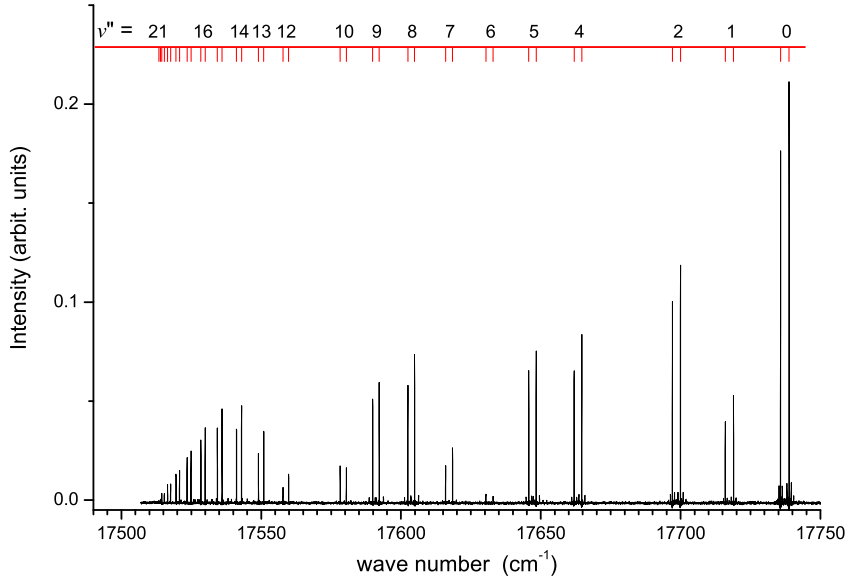,width=0.69\linewidth}
  \caption{The fluorescence progression following the excitation to the ($v'$=6, $J'$=29) rovibrational
  level in the 2$^{3}\Pi_{g}$ state. The weak lines around the strongest ones
are rotational satellites, caused by the presence of the buffer gas in the heat pipe.} \label{spectr}
\end{figure*}

The experimental setup for the production of K$_2$ molecules is
similar to that described in our previous papers \cite{Pashov:05,Docenko:06}. 
A single section heat-pipe was filled with about 10 g
of potassium (natural isotopic composition) and heated to about
600 K. Ar was used as buffer gas at a pressure of about 1-2 mbar.

For excitation of the potassium molecules we applied the laser 
lines listed in Ref.~\cite{Chu:05} and then used a Fourier-transform
spectrometer to resolve the
induced fluorescence to the a$^3\Sigma^+_{u}$ state with a typical
resolution of 0.05 cm$^{-1}$.
Two diode laser heads with an external grating cavity (DL 100 from Toptica) and 
the accompanying electronics were supplied with laser diodes delivering about
50mW at 850 nm or 100 mW at 980 nm, respectively. The lasers were superimposed 
collinearly by a dichroitic mirror and focused in the central part of the 
heat-pipe oven. The frequency of the lasers was controlled with a wavemeter
(type Highfinesse WS7), which was calibrated against the He-Ne/I$_2$ frequency 
standard in our lab in Hannover.

\subsection{Two-photon single-color excitations}

For the two-photon transitions we  applied only the 980 nm laser tuned
to the frequencies of Table II and Table III of
Ref.\cite{Chu:05}. In order to increase the detected signal
we applied a Doppler-free excitation scheme since then all molecules independent 
of their velocity classes contribute to the fluorescence intensity.
The laser beam was back reflected and refocused after its first pass through
the heat-pipe. When the laser frequency was tuned to the
center of the two-photon transition a narrow
Doppler-free peak was observed on the Doppler-broadened pedestal, which stems from
the two-photon processes by single laser beam direction. In this way we were able
to increase the intensity of the induced fluorescence by about a
factor of 5-7 and also eliminated the possible Doppler shift of the
recorded fluorescence frequencies of the 2$^3\Pi_{g}\rightarrow$~a$^3\Sigma^+_{u}$
system.

From the whole list of transitions given in \cite{Chu:05} we
registered strong discrete spectra to the \astate\ state only for
5 (10199.200 \rcm, 10226.307 \rcm, 10251.126 \rcm, 10258.240 \rcm, and 10291.740 \rcm) out of 28, which were in the wavelength region we could cover in the present experiment. The 5 used excitations gave sufficiently strong discrete fluorescence whereas the others mainly gave continuum fluorescence. We found a new two-photon excitation at 10210.991 \rcm. While scanning the laser we
frequently observed yellow-orange fluorescence also at other
frequencies, but only in few cases we were able to record
discrete spectral lines. We believe that such fluorescence comes
from bound-free transitions to the repulsive branch of the
\astate\ state. 

In some spectra (excitations at 10210.991~\rcm\ and
10199.200~\rcm) we
observed in addition to the \astate\ state fluorescence also fluorescence to the b$^{3}\Pi_{u}$ state. This could be 
helpful in a future analysis of the coupled system of b$^{3}\Pi_{u}$ and 
A$^{1}\Sigma_{u}$ states.

\subsection{Two-step two-color excitations}

The main body of experimental data comes from the two-step
excitations, which were selected from Table I of
Ref. \cite{Chu:05}. The signals in this case were much stronger
than the two-photon ones, higher levels of the 2$^{3}\Pi_{g}$ state
were excited which allowed longer progressions to the \astate\
state to be observed. A typical progression following the
excitation to the ($v'$=6, $J'$=29) level in the 2$^{3}\Pi_{g}$ state and reaching up to
$v''$=21 is shown in Fig.~\ref{spectr}.

The mutual stability of both laser frequencies with respect to each
other was somewhat critical. Therefore, we usually tuned first the 850~nm
laser to the desired transition frequency of the first step
(\Xstate - (A$^1\Sigma_{\mathrm{u}}^+\sim$ b$^3\Pi_{\mathrm{u}}$)),
then stabilized the frequency of the 980~nm laser (the second step by the transition
b$^3\Pi_{\mathrm{u}}$ - 2$^3\Pi_{\mathrm{g}}$ ) on the maximum of the
yellow-orange fluorescence appearing due to double resonance. For the 
stabilization of the laser frequency in the second step, the current of the 980 nm laser was 
modulated and the error signal was created by a Lock-in detection on 
the modulation frequency. Finally, the frequency of the 850 nm laser was fine tuned in order to maximize the 
fluorescence. During a typical scan of the Fourier spectrometer (about 20 min.) the stability of the 850 nm laser
cavity was sufficient to keep its frequency to within few tens of MHz without active stabilization. 
The second-step laser followed the slow drifts of the first one by the feedback loop
for the stabilization. This setup was sufficient to ensure stable conditions during the
recording of the Fourier-transform spectrometer.

\section{Analysis}
\label{analysis}

Initially, our identification of the observed two-photon
progressions was based on the data (transition frequencies and assignments) from Ref.~\cite{Chu:05} and the
Dunham coefficients for the a state from Ref.~\cite{Li:90}. After collecting
several clear progressions we tried to fit a potential energy
curve for the \astate\ state, but we found that the
rotational numbering N, suggested with the help of the Dunham coefficients, 
was most likely incorrect at least for one of the transitions since it turned 
out to be impossible to describe these progressions with a single
potential curve. The identification of the two-step processes in
Ref.~\cite{Chu:05} is much more reliable therefore we used it to
establish the assignment of the transitions to the \astate\ state.
With a potential curve fitted to only two such progressions (using
the pointwise potential presentation from Ref.~\cite{ipaasen}) we were able to fix the
rotational assignment also of the two-photon transitions. 

In Table~\ref{excit} we present the list of the assigned transitions and corresponding laser
frequencies used in the present two-photon and two-step excitations. Most frequencies were reported already in
Ref.~\cite{Chu:05} and  the vibrational assignment of the levels of the 2$^3\Pi_g$  state follows 
this reference, but the rotational quantum numbers are reassigned. The excitation at 10210.991 \rcm\ was detected in our study, and the vibrational numbering of the upper state is based on the Dunham coefficients of the 2$^3\Pi_g$  state reported in
Ref.~\cite{Chu:05}.

\begin{table}
\fontsize{8pt}{12pt}\selectfont
\caption{List of the assigned transitions excited by the laser frequencies used in the present 
experiment. In the first four columns the  quantum numbers for the 2$^3\Pi_g$ and
the \Xstate\ states are given, respectively. The vibrational assignment of the 2$^3\Pi_g$ 
levels is taken from Ref.\cite{Chu:05}, except for the last two-photon transition,
which was detected only in this study. In the last column the laser frequencies for the 
two-photon (one value) and the two-step excitations (two values) are listed. The uncertainties are less than 0.010 \wn}
\label{excit}
\begin{tabular}{rr|rr|r} \hline
$v'$    & $J'$  & $v''$    & $J''$  & Laser frequency ($\rcm$) \\
\hline
0	&	53	&	13	&	53	&	10199.200		\\
1	&	55	&	11	&	55	&	10291.740		\\
2	&	74	&	12	&	72	&	10258.240		\\
2	&	78	&	12	&	78	&	10251.125		\\
5	&	62	&	15	&	62	&	10226.307		\\
6	&	109	&	14	&	109	&	10210.991		\\
\hline
6	&	25	&	0	&	23	&	11641.184 + 10241.062		\\
6	&	27	&	0	&	25	&	11644.249 + 10235.844		\\
6	&	29	&	0	&	27	&	11643.657 + 10234.098		\\
6	&	31	&	0	&	29	&	11643.237 + 10231.966		\\
6	&	33	&	0	&	31	&	11642.946 + 10229.528		\\
7	&	25	&	0	&	25	&	11644.248 + 10285.797		\\
7	&	25	&	0	&	23	&	11641.185 + 10294.354		\\
7	&	27	&	0	&	25	&	11644.249 + 10289.130		\\
7	&	31	&	0	&	31	&	11642.948 + 10278.712		\\
8	&	25	&	0	&	23	&	11641.185 + 10347.338		\\
8	&	27	&	0	&	27	&	11643.657 + 10336.777		\\
8	&	29	&	0	&	27	&	11643.657 + 10340.363		\\

\hline
\end{tabular}
\end{table}

We estimated the experimental uncertainty of the Fou-rier-transform data
conservatively  to be 0.005 \rcm\ from the applied resolution of 0.05 \rcm. However, the dimensionless standard deviation
of the preliminary potential fit, being about 0.5, suggests that the primary 
uncertainty is somewhat overestimated.

\begin{figure}
  \centering
\epsfig{file=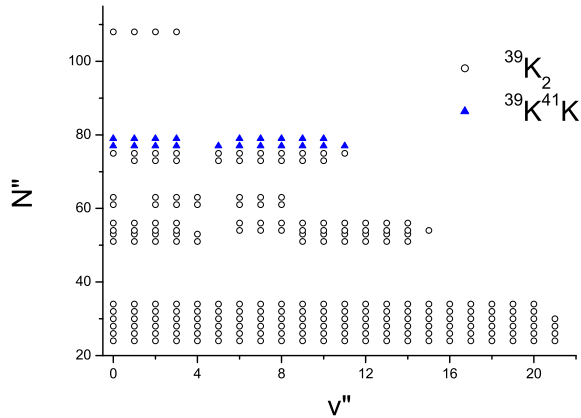,width=0.99\linewidth}
  \caption{The range of vibrational and rotational quantum numbers $v''$ and $N''$ of
the energy levels of the \astate\ state, observed in the present
study.} \label{dataset}
\end{figure}

The majority of the observed transitions was from the most abundant
isotopic combination $^{39}$K$^{39}$K. Only in one spectrum (the
two-photon excitation at 10251.125 \rcm) we found lines also from
$^{39}$K$^{41}$K. The data field of all observed rotational and
vibrational quantum numbers is given in Fig.~\ref{dataset}.  The full
list of excitation frequencies, their new assignments, and the observed progressions containing
639 transitions to 238 levels of the \astate\ state can be found in the supplementary material \cite{sup}.

\section{Coupled channels treatment}
\label{CC}
The initial pointwise potential for the \astate\ state derived in the section above is based 
only on the spectroscopic data of our experiment. As a second step 
of our analysis we included also the progressions to the \Xstate\ state 
measured by Amiot et al. \cite{Amiot:95} and fitted the complete data set to two potentials having the 
same long range behavior determined by the dispersion coefficients C$_6$, C$_8$ and C$_{10}$ and opposite exchange terms. The 
fit applied the analytic representation as described in our recent work on KRb \cite{Pashov:07}.
In order to fix the absolute position of the \astate\ state with respect to the \Xstate\ state,
we need a common origin with respect to which the energies of levels of both these states are known.
As such origins we used the term energies of the upper 2$^3\Pi$ state levels involved in 
the two-step and the two-photon processes. The transition energy to the corresponding 
\Xstate\ state levels is given by the sum of the two laser frequencies, 
whereas the transition frequencies to the \astate\ state levels were measured directly by the FTS. 

For an easy understanding and a full definition of the parameters contained in later tables 
we repeat the relevant formulas of the analytic potential representation.

The representation of the potentials is split into three regions:
the repulsive wall (R$<$R$_{inn}$), the asymptotic region (R$>$R$_{out}$),
and the intermediate region in between.
The analytic form of each potential in the intermediate range is described by a finite power expansion
with a nonlinear variable function $\xi$ of internuclear separation R:

\begin{equation}
\label{xv}
\xi(R)=\frac{R - R_m}{R + b\,R_m}
\end{equation}
\begin{equation}
\label{uanal}
\mbox{U}_{\mathrm {IR}}(R)=\sum_{i=0}^{n}a_i\,\xi(R)^i
\end{equation}

\noindent where the \{a$_i$\} are fitting parameters and $b$  and $R_m$
are chosen during the transformation process from the pointwise representation to the analytic form of equation (\ref{uanal}), 
$R_m$ is close to the value of the equilibrium separation. The potential is 
extrapolated for R $< \mbox{R}_{inn}$ with:

\begin{equation}
\label{rep}
  \mbox{U}_{\mathrm {SR}}(R)= A + B/R^{N_s}
\end{equation}

\noindent by adjusting the $A$ and $B$ parameters to get a continuous transition at $\mbox{R}_{inn}$; 
N$_s$ was 12 and 6 for \Xstate~and \astate\ states, respectively.

For large internuclear distances (R $> \mbox{R}_{out}$)
we adopted the standard long range form of molecular potentials:

\begin{equation}
\label{lrexp}
  U_{\mathrm {LR}}(R)=U_{\infty}-C_6/R^6-C_8/R^8-C_{10}/R^{10}\pm E_{\mathrm{exch}}
\end{equation}

\noindent where the exchange contribution is given by

\begin{equation}
\label{exch}
E_{\mathrm{exch}}=A_{\mathrm{ex}} R^\gamma \exp(-\beta R) 
\end{equation}
and U$_{\infty}$ set to zero for fixing the energy reference.

These potentials were applied in a coupled channels calculation including the hyperfine 
parameters and the electronic and nuclear g-factors of the potassium atoms \cite{Arimondo}
and the magnetic spin-spin coupling of the two atomic doublet states. The full hamiltonian 
was already described in several publications, e.g. in Ref. \cite{Mies:00,Laue:02}. 

The Feshbach resonances reported in \cite{Regal:03,Regal:04,Gaebler:07,Errico:07} and the two-color photoassociation
data from Ref. \cite{Wang:00} were included in the fit using the published error limits to determine the 
weighting. These data give information on asymptotic bound levels of the two isotopomers
$^{39}$K$_2$ and $^{40}$K$_2$ and, especially the Feshbach resonances, on the singlet/triplet
coupling, while the levels from photoassociation work turned out to be mainly of triplet
character. For the Feshbach resonances on $^{40}$K$_2$ we selected the results from Ref. 
\cite{Regal:04,Gaebler:07} because these are the most precise ones and should be closely
related to the two-body collision process while those from Ref. \cite {Regal:03} could be
influenced by three-body effects as studied in Ref. \cite {Smirne:07} for Rb. 

The fit was performed iteratively. First, the fit of the asymptotic levels of the photo association spectroscopy and of the magnetic fields of the Feshbach resonances varies only the lowest order dispersion term and the exchange term, keeping all other parameters fixed for a preliminary potential representation. For the second
fit step the preliminary results were used to calculate the binding energies of those levels 
to which the Feshbach resonances and the photoassociation levels correlate for the uncoupled case. 
These  calculated energies with their quantum numbers, derived directly from the calculations, were then added as data points to the data field for the full potential fit and a new fit, now for all free parameters of the three regions of each potential, was performed. 
The procedure is iterated two times to find convergence. The standard deviation of the coupled 
channels fit is $\sigma$ = 0.84 for the Feshbach resonances and the photoassociation 
data, and  $\sigma$ = 0.82  for the full potential step for both potentials together showing the 
good consistency of this approach. 

For the potential fit of the single channel case the standard deviation is 0.48 for the manifold of triplet levels 
alone and 0.82 for that of the singlet levels . At the end of the evaluation, the scattering calculations were 
extended by including d-waves for the s-wave resonances and f-waves for 
the p-wave resonance, the coupling is possible by spin-spin interaction and higher order spin-orbit interaction. The influence of the higher partial waves turned out to be insignificant with respect to the experimental uncertainty of the magnetic field determination in these cases.

The potential results are listed in Table \ref{tabX}~for the 
\Xstate\ state and in Table \ref{taba}~for the \astate\ state; the given number of digits for 
the potential parameters are not checked for their absolute need according to roundoff errors, 
but copied from the computer output. It is quite certain, that fewer digits would be 
sufficient in several cases to reproduce all observations within experimental uncertainty.

The spectroscopic data on \Xstate\ contain levels which have outer turning points up to 16.83~\AA, 
about 0.934 \wn\ below the asymptote, whereas the levels derived from the Feshbach resonances 
start at outer turning points of 27.03~\AA\ and are 0.051 \wn\ below the asymptote. This shows 
directly the remaining energy gap between the two data sets. The situation is similar 
for the \astate\ state: outer turning points from spectroscopic data up to 15.27~\AA~ 
and  from photoassociation data starting 
from 23.00~\AA, and these correspond to energies about 1.716 \wn\ and   0.136 \wn\ below the 
asymptote. At such separations the exchange energy is already negligible compared to the long range 
contribution of the dispersion terms. The small energy gaps 
of about 1~\wn~in both cases assure a reliable extrapolation to the dissociation energy. At this occasion,
one should also note that the derived dispersion coefficients closely agree to the theoretical 
values reported by Derevianko et al. \cite{Derevianko:99,Porsev:03}; deviations 
of C$_8$ and C$_{10}$ are equal within digits shown in Ref. \cite{Porsev:03} and for C$_6$ 
the present value is larger by two times of the error given in Ref. \cite{Derevianko:99} . 

The hyperfine structure of $^{39}$K$_2$ is the largest for our spectroscopic observations, 
isotopomers with $^{40}$K were not detected in our spectroscopic investigation because of low 
natural abundance. The total hyperfine structure of a single rotational state of \astate\ spans 
923 MHz with  widest spacing between adjacent levels of about 150 MHz. Thus no hyperfine 
structure could be resolved within the resolution of our spectra. We checked with coupled channels 
calculations that the general turnover from "pure" singlet/triplet character to mixed spin 
states begins for binding energies smaller than 7.0 GHz or 0.23 \wn, thus in our spectroscopic data set only 
accidental local perturbations by closely spaced singlet-triplet levels could give observable 
energy shifts of a singlet and a triplet group. We did not find any within the present data set.

\begin{table}
\fontsize{8pt}{13pt}\selectfont
\caption{Parameters of the analytic representation of the \Xstate\ state potential. The energy reference is the dissociation asymptote. Parameters with $^\ast$ are set for continuous extrapolation of the potential. }
\label{tabX}
\begin{tabular*}{1.0\columnwidth}{@{\extracolsep{\fill}}|lr|}
\hline
   \multicolumn{2}{|c|}{$R < R_\mathrm{inn}=$ 2.870 \AA}    \\
\hline
   $A^\ast$ & -0.265443197$\times 10^{4}$ \wn \\
   $B^\ast$ &  0.820372803$\times 10^{9}$  \wn \AA $^{12}$ \\
\hline
   \multicolumn{2}{|c|}{$R_\mathrm{inn} \leq R \leq R_\mathrm{out}=$ 12.000 \AA} \\
\hline
    $b$ &   $-0.40$ \\
    $R_\mathrm{m}$ & 3.92436437 \AA  \\
    $a_{0}$ &  -4450.906205 \wn\\
    $a_{1}$ &  0.70355350020116$$ \wn\\
    $a_{2}$ &  0.13671174694653$\times 10^{5}$ \wn\\
    $a_{3}$ &  0.10750698806556$\times 10^{5}$ \wn\\
    $a_{4}$ & -0.20932329414778$\times 10^{4}$ \wn\\
    $a_{5}$ & -0.19384823376156$\times 10^{5}$ \wn\\
    $a_{6}$ & -0.49209429682855$\times 10^{5}$ \wn\\
    $a_{7}$ &  0.11026750296026$\times 10^{6}$ \wn\\
    $a_{8}$ &  0.72867383247088$\times 10^{6}$ \wn\\
    $a_{9}$ & -0.29310771189374$\times 10^{7}$ \wn\\
   $a_{10}$ & -0.12407064957537$\times 10^{8}$ \wn\\
   $a_{11}$ &  0.40333954923169$\times 10^{8}$ \wn\\
   $a_{12}$ &  0.13229846082365$\times 10^{9}$ \wn\\
   $a_{13}$ & -0.37617672560621$\times 10^{9}$ \wn\\
   $a_{14}$ & -0.95250412147591$\times 10^{9}$ \wn\\
   $a_{15}$ &  0.24655585672079$\times 10^{10}$ \wn\\
   $a_{16}$ &  0.47848258035225$\times 10^{10}$ \wn\\
   $a_{17}$ & -0.11582132128030$\times 10^{11}$ \wn\\
   $a_{18}$ & -0.17022518278642$\times 10^{11}$ \wn\\
   $a_{19}$ &  0.39469335089283$\times 10^{11}$ \wn\\
   $a_{20}$ &  0.43141949807984$\times 10^{11}$ \wn\\
   $a_{21}$ & -0.97616955371081$\times 10^{11}$ \wn\\
   $a_{22}$ & -0.77417530660299$\times 10^{11}$ \wn\\
   $a_{23}$ &  0.17314133620597$\times 10^{12}$ \wn\\
   $a_{24}$ &  0.96118849014390$\times 10^{11}$ \wn\\
   $a_{25}$ & -0.21425463052972$\times 10^{12}$ \wn\\
   $a_{26}$ & -0.78513081744374$\times 10^{11}$ \wn\\
   $a_{27}$ &  0.17539493137145$\times 10^{12}$ \wn\\
   $a_{28}$ &  0.37939637130987$\times 10^{11}$ \wn\\
   $a_{29}$ & -0.85271868544557$\times 10^{11}$ \wn\\
   $a_{30}$ & -0.82123528497789$\times 10^{10}$ \wn\\
   $a_{31}$ &  0.18626451763727$\times 10^{11}$ \wn\\
\hline
   \multicolumn{2}{|c|}{$R_\mathrm{out} < R$}\\
\hline
  ${U_\infty}$ & 0.0 \wn    \\
 ${C_6}$ &    0.1889676057$\times 10^{8}$ \wn\AA$^6$      \\
 ${C_{8}}$ &  0.5527948928$\times 10^{9}$ \wn\AA$^8$   \\
 ${C_{10}}$ & 0.2185553504$\times 10^{11}$ \wn\AA$^{10}$   \\
 ${A_{ex}}$ & 0.21698263$\times 10^{5}$ \wn\AA$^{-\gamma}$   \\
 ${\gamma}$ & 5.19500    \\
 ${\beta}$ & 2.13539 \AA$^{-1}$   \\
\hline
 \multicolumn{2}{|c|}{Derived constants:} \\
\hline
\multicolumn{2}{|l|}{equilibrium distance:\hspace{2.2cm} $R_e^X$= 3.92436(5) \AA} \\
\multicolumn{2}{|l|}{electronic term energy:\hspace{1.6cm} $T_e^X$= -4450.906(50) \wn}\\
\hline
\end{tabular*}
\end{table}

\begin{table}
\fontsize{8pt}{13pt}\selectfont
\caption{Parameters of the analytic representation of the \astate\ state potential. The energy reference is the dissociation asymptote. Parameters with $^\ast$ are set for continuous extrapolation of the potential.  }
\label{taba}
\begin{tabular*}{1.0\columnwidth}{@{\extracolsep{\fill}}|lr|}
\hline
   \multicolumn{2}{|c|}{$R < R_\mathrm{inn}=$ 4.750 \AA}    \\
\hline
   $A^\ast$ & -0.559417167$\times 10^{3}$ \wn \\
   $B^\ast$ & 0.6432888245$\times 10^{7}$  \wn \AA $^{6}$ \\
\hline
   \multicolumn{2}{|c|}{$R_\mathrm{inn} \leq R \leq R_\mathrm{out}=$ 12.000 \AA}    \\
\hline
    $b$ &   $-0.300$              \\
    $R_\mathrm{m}$ & 5.73392370 \AA  \\
    $a_{0}$ &  -255.016965 \wn\\
    $a_{1}$ & -0.44746842073489$$ \wn\\
    $a_{2}$ &  0.20951803151410$\times 10^{4}$ \wn\\
    $a_{3}$ & -0.17131183698021$\times 10^{4}$ \wn\\
    $a_{4}$ & -0.17772657861768$\times 10^{4}$ \wn\\
    $a_{5}$ &  0.29413668239428$\times 10^{4}$ \wn\\
    $a_{6}$ & -0.20171041930434$\times 10^{5}$ \wn\\
    $a_{7}$ & -0.35711976066048$\times 10^{5}$ \wn\\
    $a_{8}$ &  0.59856336996119$\times 10^{6}$ \wn\\
    $a_{9}$ & -0.71043946542935$\times 10^{6}$ \wn\\
   $a_{10}$ & -0.61713401161663$\times 10^{7}$ \wn\\
   $a_{11}$ &  0.19365677976135$\times 10^{8}$ \wn\\
   $a_{12}$ &  0.67930464983208$\times 10^{7}$ \wn\\
   $a_{13}$ & -0.12020038974090$\times 10^{9}$ \wn\\
   $a_{14}$ &  0.21603950703685$\times 10^{9}$ \wn\\
   $a_{15}$ & -0.63530871042880$\times 10^{8}$ \wn\\
   $a_{16}$ & -0.52391336483017$\times 10^{9}$ \wn\\
   $a_{17}$ &  0.15913325190081$\times 10^{10}$ \wn\\
   $a_{18}$ & -0.24792577649852$\times 10^{10}$ \wn\\
   $a_{19}$ &  0.20325982754798$\times 10^{10}$ \wn\\
   $a_{20}$ & -0.68043793785293$\times 10^{9}$ \wn\\
\hline
   \multicolumn{2}{|c|}{$R_\mathrm{out} < R$}\\
\hline
  ${U_\infty}$ & 0.0 \wn    \\
 ${C_6}$ &    0.1889676057$\times 10^{8}$ \wn\AA$^6$      \\
 ${C_{8}}$ &  0.5527948928$\times 10^{9}$ \wn\AA$^8$   \\
 ${C_{10}}$ & 0.2185553504$\times 10^{11}$ \wn\AA$^{10}$   \\
 ${A_{ex}}$ &-0.21698263$\times 10^{5}$ \wn\AA$^{-\gamma}$   \\
 ${\gamma}$ & 5.19500    \\
 ${\beta}$ & 2.13539 \AA$^{-1}$   \\
\hline
 \multicolumn{2}{|c|}{Derived constants:} \\
\hline
\multicolumn{2}{|l|}{equilibrium distance:\hspace{2.2cm} $R_e^a$= 5.7344(1) \AA} \\
\multicolumn{2}{|l|}{electronic term energy:\hspace{1.6cm} $T_e^a$= -255.017(50) \wn}\\
\hline
\end{tabular*}
\end{table}

\section{Discussion and conclusion}
\subsection{Potentials and dissociation energies}
The potentials determined in the present work describe the spectroscopic observation and the results 
from cold collisions within the experimental accuracy.
Only the very few data obtained  by \cite{Li:90,Zhao:96} from fluorescence progressions
using a grating spectrometer show deviations beyond the reported accuracy. The standard 
deviation of these series derived with the help of the new potentials are 0.70 \wn~and 0.37~\wn, respectively, 
while the reported experimental accuracies are  0.17~\wn\ and 0.05~\wn . We get an averaged 
shift between the two series of these reports of 2.54~\wn~which is close to the shift derived 
in \cite{Zhao:96} and interpreted in that paper as a calibration difference in both grating 
instruments. Thus only the unusual large scatter in both series remains unexplained. Trials of
reassignment of these spectra in N and v quantum numbers remained unsuccessful. 

From the potentials of the ground state one can read off the dissociation energies D$_e$, 4450.906 (50) \wn\ for the 
\Xstate\ state and 255.017 (50) \wn\ for the \astate\ state. Zhao et al.\cite{Zhao:96} reported values for these states 
as 4450.674(72) \wn\ and 252.74(12) \wn, respectively. For the \Xstate\ state both values almost 
agree, but for the \astate\ state a clear discrepancy is found, which is certainly related to the 
calibration problem and the surprisingly big scatter of the results from the grating spectrographs. Because of the new large body of data on the \astate~state with an accuracy of better than 0.005~\wn~we recommend undoubtedly the 
application of the new result. Because the derivation of the position of the potential minimum is 
dependent on the mathematical representation of the potential curve in principle and 
this dependence might show up at the present level of accuracy, we prefer to give the dissociation
energy with respect to an observable bound level, e.g. v=0, J=0, normally named by D$_0$, but 
this value is then isotope dependent. For the main isotopomer $^{39}$K$_2$ we obtain D$_0= 
4404.816(50)$~\wn\ for the singlet state and D$_0=244.523(50)$~\wn\ for the triplet state, 
these values are better suited for comparing results of future studies of expected high level of accuracy. 

Recently, high resolution molecular beam spectroscopy of asymptotic levels of the state A$^1\Sigma^+_u$ were reported by our group \cite{Falke:06}. With the help of these data a very reliable value of D$_0$ of the \Xstate~state of the main isotopomer $^{39}$K$_2$ was derived, namely 4404.808(4) \wn, which agrees with the new value above. But this value is an order of magnitude more precise than the present, completely independently derived value. The good agreement between both experimental results certainly confirms our conclusion drawn in the paragraph above. Furthermore, incorporating the precise dissociation energy as a data point of the level v=0 and J=0 of \Xstate\ with respect to the dissociation limit in the fit, it directs us to reduce the error limit of the dissociation energy of the state \astate\ significantly, because we measured the relative position of the triplet and singlet level scheme by our two-photon and two-step investigation, as given in Tab. \ref{excit}. This results to D$_e= 255.017(10)$~\wn\ or D$_0=244.523(10)$~\wn\ of $^{39}$K$_2$. 

\subsection{Cold collisions and Feshbach resonances}
\begin{figure}
  \centering
\epsfig{file=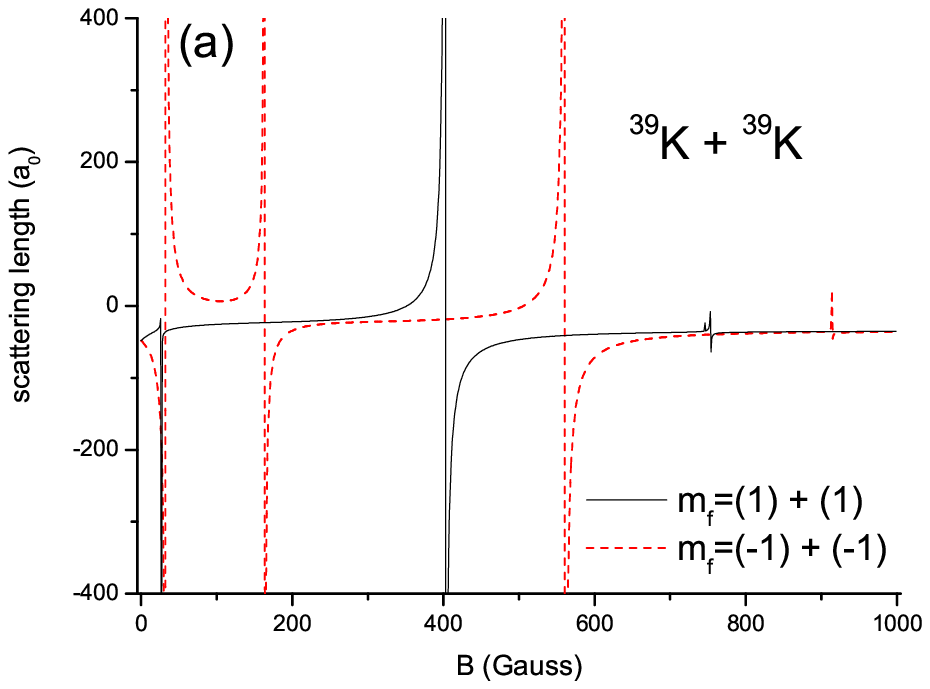,width=0.99\linewidth}
\epsfig{file=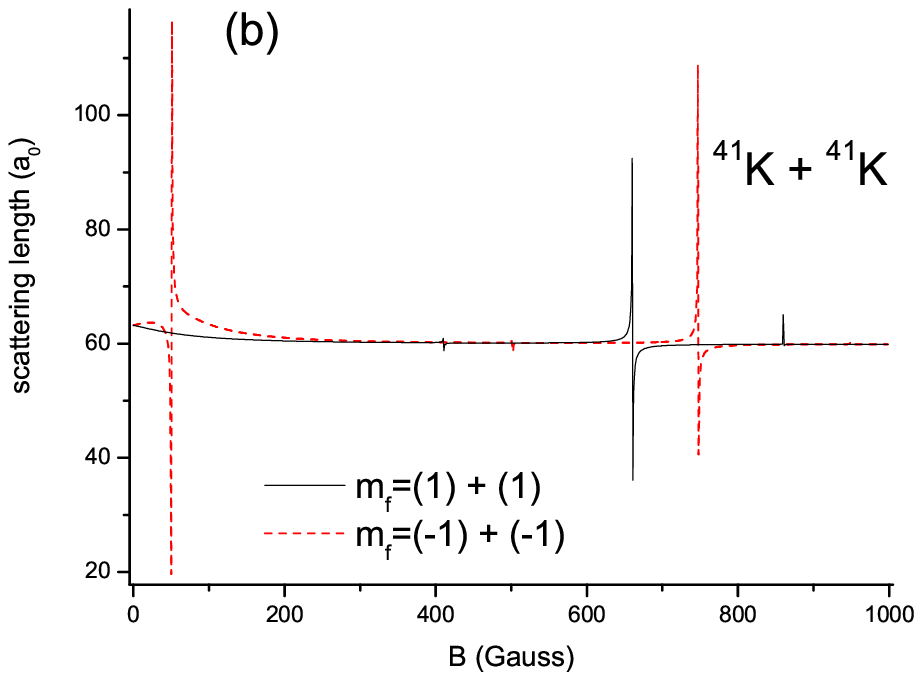,width=0.99\linewidth}
\epsfig{file=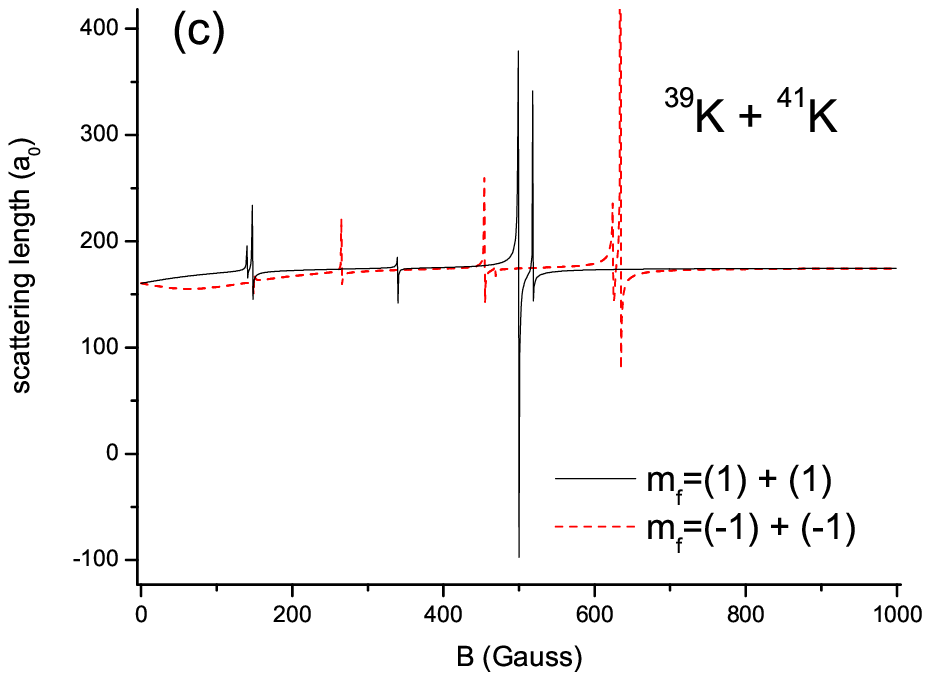,width=0.99\linewidth}
  \caption{s-wave Feshbach resonances of (a) $^{39}$K, (b)$^{41}$K and (c) their combination. For all cases the atomic  
angular momentum is f = 1, the projection m$_f$ on the space fixed axis is given in each graph. The unit of the scattering length is the Bohr radius $a_0 = 0.5292 \times 10^{-10}m$} \label{resonances}
\end{figure}

In the data evaluation three different isotopomers of potassium are included, namely $^{39}$K$_2$,
$^{40}$K$_2$, and $^{39}$K$^{41}$K. Thus, it might be possible to get a first answer, if mass 
scaling is applicable in the case of potassium at the present level of accuracy. 
At a first glance the obtained standard deviations are below 1.0, see section \ref{CC}. 
Thus the evaluation is within the reported experimental accuracies. However, looking more closely to 
the deviations of the highly precise Feshbach resonances, where a magnetic field uncertainty of 
0.05 G relates to an uncertainty in the order of 100 kHz in the frequency scale, the fit for 
$^{40}$K$_2$ is excellent, but the scatter between the deviations for $^{39}$K$_2$ is  fairly large and for such small set of data too often at the limit of the experimental accuracy. This is not so obvious in the published fit of Ref. \cite{Errico:07}, because they used the less precise data of $^{40}$K from Ref.\cite{Regal:03} in their fit. We also would like to note, that the standard deviation, given as reduced $\chi ^2$ with the value 0.52 is probably in error, we obtained with their data 1.04. Also for some data of the few photoassociation measurements on the isotope $^{39}$K the deviations come close to the reported experimental uncertainties. Thus we 
would like to recommend new experiments for getting improved results from two-color 
photoassociation, which are presently reported with a 40 MHz uncertainty limit, and to extend 
the measurements of Feshbach resonances, for which we will make
predictions below (Fig. \ref{resonances}). In the same spirit we prepare presently high precision
molecular beam studies on potassium as we performed in the case of Na$_2$ \cite{Elbs:99,Samuelis:01} 
some years ago. All this together might give the proper limit for mass scaling, 
i.e. of the Born-Oppenheimer approximation, for the ground states of potassium atom pairs. For
the excited asymptote s + p we already reported an experimental evidence of necessary corrections
to the Born-Oppenheimer approximation \cite{Falke:07}.

Assuming the validity of the Born-Oppenheimer approximation or i.e. mass scaling, the potentials
allow reliable calculations of scattering lengths of the full manifold of isotopomers. The results 
are given in Table \ref{length} along with the maximum vibrational quantum number 
within the potentials for the lowest rotational state J=N=0. These results agree with the latest 
determinations from cold collision experiments; references are cited in the appropriate isotopomer column, where experimental data 
were directly used for the derivation of that isotopomer. Other predictions exist in the literature, which are close to the 
ones given in Table \ref{length}. The predictions of Table  \ref{length} are homogeneous, 
because they all are derived from the same potential model. The slight difference for the triplet scattering length of $^{39}$K between Ref. \cite{Errico:07} and the present value originates from differences in the magnitude of the exchange force \cite{Simoni:07} used in both approaches. The combination of spectroscopic and Feshbach resonance data results in the increased value as given in the potential tables above.

Ultracold potassium ensembles are often used for modeling condensed matter physics or
cooling processes in connection with other species. To guide new experiments we calculated Feshbach
resonances for the species $^{39}$K, $^{41}$K and their combination at the lowest atomic asymptote 
m$_f$ = 1 + 1 and at the low field seeking asymptote within a MOT at m$_f$= (-1) + (-1). The 
results are collected in Figure \ref{resonances} and show very promising structures at fairly 
low fields, which are of easy access by experiments. The calculations were done with a step 
size of 1 Gauss and thus in the cases of sharp resonances the curves are not going up to $\pm \infty$. Additionally, 
the scale of the vertical axis in Fig. \ref{resonances} does not extend to very large positive and negative scattering lengths; instead it is chosen to illustrate the behavior of the scattering length in the region of the bottom of the resonance profile, which is important for fine tuning of the two-boy interaction for experiments. In Ref. \cite{Errico:07} similar predictions are reported which are mainly consistent with ours. The reader should note that figure 5 and 6 are interchanged in \cite{Errico:07}.

In Fig. \ref{resonances} (a) the broad resonances at about 400 G was used by Roati et al. 
\cite{Roati:07} to obtain Bose-Einstein condensation for $^{39}$K. For the two homonuclear 
cases calculated resonances at low field around 40 to 50 G appear, which allow the tuning of 
the two-body interaction in convenient field ranges. The resonance structure in the heteronuclear
case is especially rich and would allow a very careful study of the validity of mass scaling. 
Fig. \ref{resonances} gives only examples, but we present in this paper all information needed 
for further calculations of collision properties at different atomic asymptotes. From the present fitting results we conclude 
that predictions of Feshbach resonances with our model potentials should be accurate to better 
than 1 Gauss. For the effective spin-spin coupling only the magnetic dipole-dipole contribution 
of atomic pairs was needed in the analysis by \cite {Ticknor:04} for the splitting of the p-wave 
resonance in $^{40}$K. Further studies on such resonances or two-color photoassociation 
spectroscopy with improved resolution could yield the missing information for deriving the second 
order spin-orbit contribution to the effective spin-spin coupling as it was obtained for Na$_2$ 
by de Araujo et al. \cite{Fatemi:03}, giving further improvement on the prediction of collision 
properties.

\begin{table}
\fontsize{7pt}{12pt}\selectfont
\caption{Scattering lengths (unit $a_0=0.5292$ \AA ) and maximum vibrational quantum numbers within each potential for different isotopomers of potassium. }
\label{length}
\begin{tabular}{r|rr|rr|rr} \hline
isotope & \multicolumn{2}{c|}{$a_{singlet}$} & \multicolumn{2}{c|}{$a_{triplet}$} & \multicolumn{2}{c}{ $v_{max}$} \\
& others & present & others & present & singlet & triplet  \\
\hline
    $39/39$ &   $138.90(15)$\cite{Errico:07}    &   $ 138.85$ &   $-33.3(3)$\cite{Errico:07} &   $ -33.15 $ &   $85$ &   $ 26 $   \\
    $39/40$ &   $  $    &   $ -2.53$ &   $  $ &   $ -1926 $ &   $85$ &   $ 26 $   \\
    $39/41$ &   $  $    &   $ 113.16$ &   $  $ &   $ 177.1 $ &   $86$ &   $ 27 $   \\
    $40/40$ &   $104.8 (4)$\cite{Loftus:02}    &   $ 104.45$ &   $174 (7)$\cite{Loftus:02} &   $ 169.6 $ &   $86$ &   $ 27 $   \\
    $40/41$ &   $  $    &   $ -54.17 $ &   $  $ &   $ 97.26 $ &   $86$ &   $ 27 $   \\
    $41/41$ &   $  $ &   $ 85.43 $ &   $78 (20)$   \cite{Modugno:01} &   $ 60.35 $ &   $87$ &   $ 27 $   \\
\hline
\end{tabular}
\end{table}

\subsection {Summary}
From high resolution Fourier-transform spectroscopy new spectroscopic information is obtained for the  \astate\ state of K$_2$. It is combined with results from most recent cold collision studies \cite{Regal:04,Gaebler:07,Errico:07} and photoassociation spectroscopy \cite{Wang:00} by other laboratories and with previous spectroscopic results on the \Xstate\ state \cite{Amiot:95} to obtain potential curves for the coupled system ( \Xstate\ - \astate\ ). From this overall homogeneous approach of the derived potentials in connection with atomic hyperfine parameters and magnetic g-factors \cite{Arimondo} ultracold collisions are reliably modeled. Corrections to the Born-Oppenheimer approximation or the so called mass scaling are not yet seen within experimental accuracy. New measurements are proposed from which a new limit on the validity of mass scaling could be derived and a deviation might become obvious. Furthermore, new highly resolved measurements of deeply bound triplet states will give important information with which the assumption of using atomic parameters for describing the molecular hyperfine splitting can be checked.

\section{Acknowledgments}

The work is supported by DFG through SFB 407 and GRK 665. A.P.
acknowledges a partial support from the Bulgarian National Science
Fund Grants MUF 1506/05 and VUF 202/06 and from Sofia University through grants 72/2006 and 21/2007.


\end{document}